\documentclass[a4paper,11pt]{article}
\pdfoutput=1 
\usepackage{authblk}
\usepackage[T1]{fontenc}
\usepackage{graphicx}
\usepackage{babel}
\usepackage{amsmath,amssymb,xcolor}
\pdfoutput=1 
\usepackage{jheppub}
\usepackage{blindtext}
\usepackage{hyperref}
\usepackage{xspace}
\allowdisplaybreaks

\usepackage[normalem]{ulem}
\usepackage{multirow}
\usepackage{bashful}
\usepackage{todonotes}
\usepackage{caption}
\usepackage{subcaption}
\usepackage{comment}

\newcommand{\LiteRed}{\texttt{LiteRed}\xspace}
\newcommand{\Libra}{\texttt{Libra}\xspace}

\newcommand{\e}{\ensuremath{\epsilon}}

\newcommand{\inlinegraphics}[2][]{\ensuremath{\vcenter{\hbox{\includegraphics[#1]{#2}}}}}

\title
{Master  integrals for $e^{+}e^{-}\rightarrow2\gamma$ process at large energies and angles.}
\author[1]{Roman N. Lee,}
\author[1, 2]{ Vyacheslav A. Stotsky}
\affiliation[1]{Budker Institute of Nuclear Physics, 630090, Novosibirsk, Russia}
\affiliation[2]{Novosibirsk State University, 630090, Novosibirsk, Russia}
\emailAdd{r.n.lee@inp.nsk.su}
\emailAdd{stotsky.slava@gmail.com}

\abstract{
	We calculate two-loop massive master integrals for $e^{+}e^{-}\rightarrow2\gamma$ in terms of generalized power series with respect to electron mass. The coefficients of this series are expressed via Goncharov's polylogarithms. Our approach exploits a number of modern multiloop methods: IBP reduction, differential equations for master integrals, Frobenius method, reduction to $\e$-form, and DRA method.
}

\begin{document}
	\maketitle
	\flushbottom

	\section{Introduction}

	The leading and next-to-leading order differential cross sections of the basic QED processes have been calculated already in the second half of the 20th century. However, the ongoing and future collider experiments require the next-to-next-to-leading order (NNLO) theoretical results. Fortunately, modern techniques of multiloop calculations can meet these needs.

	Note that the exact calculation of the two-loop diagrams with massive internal lines is notoriously much more complicated than the corresponding massless diagrams. In electron-positron colliders, the beam energy and the characteristic momentum transfer are much larger than the electron mass which can therefore be considered as a small parameter. However, this mass cannot be completely neglected due to collinear divergences in massless amplitudes. When the small mass is taken into account, these divergences transform into mass logarithms.

	In order to calculate the massive QED differential cross sections one can try the approach based on the factorization formula \cite{becher2007two}, which connects massive and massless amplitudes. This approach is claimed to reconstruct the massive amplitude up to power corrections. It was used in many calculations, including those for $e \mu \rightarrow e \mu$ \cite{Broggio:2022htr}, $e^{+} e^{-} \rightarrow \gamma \gamma^{*}$ \cite{fadin2023two}, and $e^{+}e^{-} \rightarrow 2\gamma$ \cite{naterop2021electron} processes.

	The validity of the mass factorization formula may, in principle, be questioned as it is deeply based on the assumption that only soft, hard and collinear regions contribute to the amplitude \cite{becher2007two}. A more direct approach is to use the Frobenius method and develop the generalized series expansion in electron mass. Recent work \cite{delto2023two} on Bhabha and M\o{}ller scattering uses this approach.

	In the present paper, we calculate the two-loop massive master integrals for the process $e^{+}e^{-}\rightarrow2\gamma$ in the high-energy large-angle approximation
	treating the electron mass $m$ as a small parameter. More specifically, we consider the kinematic region $m^{2} \ll s,|t|,|u|$, where
	\begin{align}
		s=(p_1+p_2)^2, \quad t=&(p_1-k_1)^2, \quad u=(p_1 - k_2)^2=2m^2 - s - t, \nonumber\\
		&s > 0, \,\,\,\,\, t,u<0.
	\end{align}
	Here $p_1$ and $p_2$ are the momenta of the initial electron and positron respectively, $k_1$ and $k_2$ are the momenta of the final photons.
	We use the Frobenius method and obtain the result in the form of generalized power series with respect to $m$ with coefficients expressed via Goncharov's polylogarithms of argument $\tau=-t/s$ with indices $0$ and $1$.

	\section{Details of the calculation}
	We set $s=1$ and use dimensional regularization, $d=4-2\epsilon$.
	Let us list the main stages of our calculation.
	\begin{itemize}
		\item First, we find IBP reduction rules exact in $m^2$ and construct differential systems for master integrals with respect to $m^2$ and $\tau=-t$.
		\item We reduce the first system to Fuchsian form at the point $m^2 = 0$ and construct the general solution in terms of generalized power series.
		\item In order to fix the special solution, we need to fix sufficiently many asymptotic coefficients in small-mass asymptotics. These coefficients are nontrivial functions of $\tau$ and are difficult to find by a direct integration.
		\item Therefore we obtain the differential system with respect to $\tau$ for these coefficients and reduce it to $\e$-form. Then we find a general solution for this system in terms of Goncharov's polylogarithms.
		\item In order to fix the boundary conditions, we need to calculate sufficiently many asymptotic coefficients in the double asymptotics $m^2\to 0$, then $\tau \to 0$. These coefficients depend on $d=4-2\e$ only, however their direct calculation using the expansion by regions is still a difficult problem.
		\item Therefore, we obtain a dimensional recurrence for those coefficients and solve it using the DRA method  \cite{tarasov1996connection,lee2010space}. In order to obtain the necessary information about the analytical properties of these coefficients, we derive and use their Mellin-Barnes representation.
	\end{itemize}

	\subsection{IBP reduction}
\begin{figure}
		\begin{subfigure}{\textwidth}
			\includegraphics[width=\textwidth]{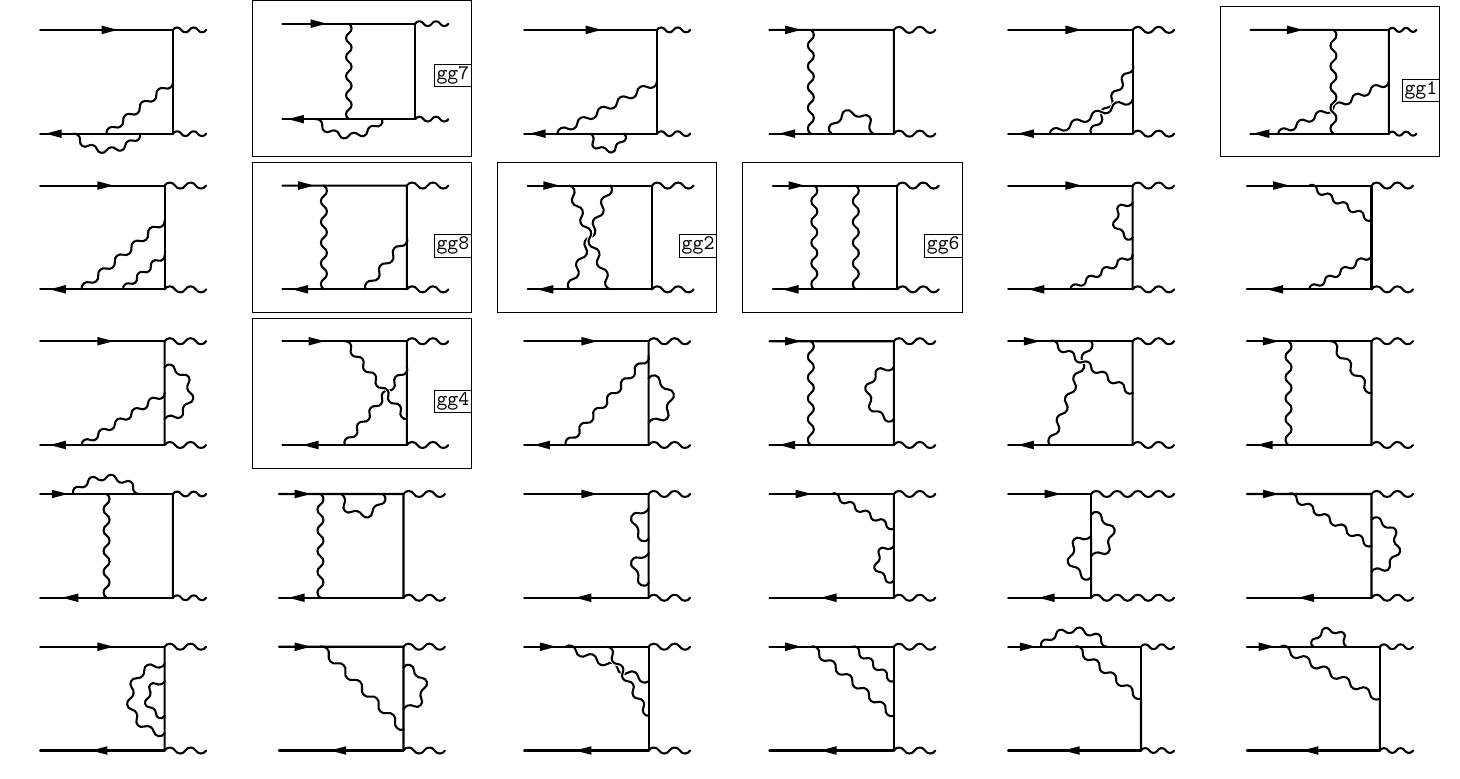}
			\caption{Pure photonic diagrams.}
			\label{fig:ph}
		\end{subfigure}
		\begin{subfigure}{\textwidth}
			\centering\includegraphics[width=0.66\textwidth]{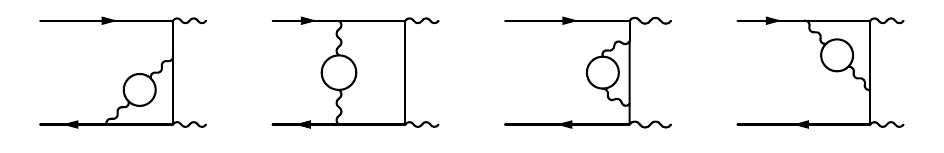}
			\caption{Diagrams with insertion of polarization operator.}
			\label{fig:po}
		\end{subfigure}
		\begin{subfigure}{\textwidth}
			\centering\includegraphics[width=0.44\textwidth]{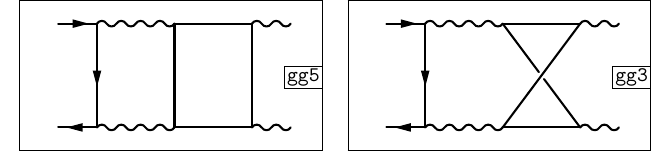}
			\caption{Diagrams with insertion of light-by-light block.}
			\label{fig:lbl}
		\end{subfigure}
		\caption{Two-loop diagrams for the process $e^+ e^- \rightarrow 2\gamma$.}
		\label{fig:diagrams}
	\end{figure}

	The amplitude of the process is given by the sum of the diagrams\footnote{Each diagram allows for two different labelings of external photon legs.} in Fig. \ref{fig:diagrams}. All diagrams are expressed in terms of the integrals of the families $\texttt{gg}_1,...,\texttt{gg}_8$ defined via
	\begin{equation}
		\texttt{gg}_i(n_1,..,n_9|\tau,m^2,d)=e^{2\e\gamma}\int_{\mathbb{R}^d}\frac{d^{d}l_1 \cdot d^{d}l_2}{(i \pi)^{\frac{d}{2}}(i \pi)^{\frac{d}{2}}}\cdot \frac{D_8 ^{-n_8} \cdot D_9 ^{-n_9}}{D_1 ^{n_1} \cdot D_2 ^{n_2} \cdots D_7 ^{n_7}}, \quad n_i \in \mathbb{Z} \label{bases}
	\end{equation}
	where $\gamma=0.577...$ is the Euler constant and $D_1,\dots,D_7$ are the denominators in the corresponding framed diagrams in Fig. \ref{fig:diagrams} while $D_{8}$, $D_{9}$ are irreducible numerators. The remaining diagrams can be expressed in terms of the same families with the replacement $\tau\to 1-\tau$.

	We discover 165 master integrals and find the IBP reduction rules exactly in the electron mass. We attach the file with the definitions of the families and master integrals as a separate file (see Section \ref{sec:files}).
	\subsection{Differential equations and dimensional recurrence relations}
	We obtain two differential systems with respect to $m^2$ and $\tau$ and  dimensional recurrence relations for master integrals $\boldsymbol{j}=\boldsymbol{j}(m^2,\tau,\epsilon)$:
	\begin{subequations}
		\label{de1}
		\begin{align}
			\partial_{m^2} \boldsymbol{j} &=M_{m} (m^2,\tau,\epsilon) \, \boldsymbol{j}, \label{dem1}\\
			\partial_{\tau} \boldsymbol{j} &= M_{\tau} (m^2,\tau,\epsilon) \, \boldsymbol{j},  \label{det1}\\
			\boldsymbol{j}(m^2,\tau,\epsilon+1) &=M_{\e}(m^2,\tau,\epsilon) \, \boldsymbol{j}(m^2,\tau,\epsilon), \label{drr0}
		\end{align}
	\end{subequations}
	where $M_{m}$, $M_{\tau}$, $M_{\e}$ are rational matrices. The differential systems cannot be reduced to $\epsilon$-form. Instead, we will use the approach of Ref.  \cite{lee2018solving} based on the Frobenius method and construct a general solution of \eqref{dem1} in terms of generalized power series in $m^2$.

	According to Ref. \cite{lee2018solving}, we first find the transformation \footnote{We use \Libra \cite{lee2021libra} for differential system reduction.}
	\begin{equation}
		\boldsymbol{j}=T_{m}(m^2,\tau,\e) \boldsymbol{J}  \label{transf1}
	\end{equation}
	to reduce the Poincar\'e rank of the differential system \eqref{dem1} at $m^2 =0$ to $0$ and to get rid of resonances. Thus, we have
	\begin{equation}
		\partial_{m^2} \boldsymbol{J} =\widetilde{M} \boldsymbol{J}, \label{de2}
	\end{equation}
	where
	\begin{equation}
		\widetilde{M}=T^{-1}(M_{m}T-\partial_{m^2}T) = \frac{\widetilde{M}_{-1} (\tau,\epsilon)}{m^2} + \sum_{n=0}^{\infty} \widetilde{M}_{n}  m^{2n} \label{de22}
	\end{equation}
	and $\widetilde{M}_{n} =\widetilde{M}_{n} (\tau,\epsilon)$ are rational matrices. The solution of \eqref{de2} can be represented in the form
	\begin{equation}
		\boldsymbol{J}=U_{m}(m^2,\tau,\epsilon) \cdot L_{m} (\tau,\epsilon) \cdot \boldsymbol{c}(\tau,\epsilon), \label{de3} \\
	\end{equation}
	where
	\begin{equation}
		U_{m}(m^2,\tau,\epsilon)  =  \sum_{  i}(m^2)^{\lambda_{i}}\sum_{n=0}^{\infty}\sum_{k=0}^{K_{i}}A_{n+\lambda_{i},k} (\tau,\e) \cdot m^{2n} \cdot \frac{ \ln^{k} m^2}{k!}. \label{de31}
	\end{equation}
	Here  $A_{n+\lambda_{i},k}$ are the matrix coefficients, $\lambda_{i}$ are the eigenvalues of $\widetilde{M}_{-1}$ and $K_i$ are pole multiplicity minus one of the resolvent $R_{\lambda} = \frac{1}{\lambda-\widetilde{M}_{-1}}$ at $\lambda = \lambda_{i}$. From the explicit form of $\widetilde{M}_{-1}$ we have
	\begin{align}
		\{\lambda_{1},\,\lambda_{2},\,\dots,\,\lambda_{7} \}  & = \{0,-\frac{1}{2}  -  3\epsilon,-\frac{1}{2}-2\epsilon,-4\epsilon,-3\epsilon,-2\epsilon,-\epsilon\}, \\
		\{K_{1},K_{2},\dots,K_{7} \} & =  \{0,0,0,0,0,2,1\}.
	\end{align}
	The ``adapter'' matrix $L_{m} (\tau,\epsilon)$ in Eq. \eqref{de3} relates the boundary constants and the coefficients $\boldsymbol{c}(\tau,\epsilon)$ in the asymptotic expansion of the master integrals at $m^2\to 0$. Note that \Libra allows one to pick the set of the asymptotic coefficients $\boldsymbol{c}$ and to find the corresponding matrix $L_m$ automatically.

	The leading coefficients $A_{\lambda_{i},k}$ are fixed from the identity
	\begin{equation}
		\sum_{  i}(m^2)^{\lambda_{i}}\sum_{k=0}^{K_{i}}A_{\lambda_{i},k} (\tau,\e) \cdot \frac{ \ln^{k} m^2}{k!} = (m^2)^{\widetilde{M}_{-1}}\,.
	\end{equation}
	Substituting \eqref{de3} into \eqref{de2} we get the recurrence relations for the coefficients $A_{n+\lambda,k}$ with $n>0$:
	\begin{equation}
		A_{n+\lambda,k}=(n+\lambda-\widetilde{M}_{-1})^{-1} \biggr[ \sum_{n'=0}^{n-1} \widetilde{M}_{n'} A_{n+\lambda-n'-1,k}-A_{n+\lambda,k+1} \biggr]. \label{de4}
	\end{equation}
	Thus, we can calculate $A_{n+\lambda,k}$ if the terms $\widetilde{M}_{n'}$ with $n' < n$ of the expansion \eqref{de22} are known. In this work, we obtain $U_m$ with the accuracy $O(m^{12})$ and exactly in $\tau$ and $\epsilon$.

	To find the asymptotic coefficients $\boldsymbol{c} (\tau,\epsilon)$, we construct the differential equations and dimensional recurrence relations for them by substituting \eqref{de3} and \eqref{transf1} into \eqref{det1} and \eqref{drr0}:
	\begin{subequations}
		\begin{align}
			\partial_{\tau} \boldsymbol{c} (\tau,\epsilon) & = S(\tau,\epsilon)   \boldsymbol{c} (\tau,\epsilon), \label{dec1} \\
			\boldsymbol{c}(\tau,\epsilon+1) &= W (\tau,\epsilon)  \boldsymbol{c}(\tau,\epsilon), \label{drr1}
		\end{align}
	\end{subequations}
	where
	\begin{subequations}
		\begin{align}
			S(\tau,\e)  &= (T_{m} U_{m} L_{m})^{-1}  \bigr[ M_{\tau}T_{m}U_{m}L_{m} - \partial_{\tau} (T_{m}U_{m}L_{m}) \bigr] , \label{dec11} \\
			W(\tau,\e) &=  [{T}_{m} U_m L_m]_{\epsilon\rightarrow\epsilon+1}^{-1} \,\,  M_{\e}   \,  \, [{T}_{m} U_m L_m]. \label{drr11}
		\end{align}
	\end{subequations}

	Note that the left-hand side of \eqref{dec1} and \eqref{drr1} does not depend on $m^2$.

	Therefore, we can obtain $S(\tau,\e)$ and $W(\tau,\e)$  expanding the right-hand side of \eqref{dec11} and \eqref{drr11} up to a sufficiently high order.

	Using the transformation
	\begin{equation}
		\boldsymbol{c}=T_{\tau} (\tau,\e)  \tilde{\boldsymbol{c}} , \label{trnsf2}
	\end{equation}
	we reduce the system \eqref{dec1} to $\epsilon$-form
	\begin{equation}
		\partial_{\tau} \tilde{\boldsymbol{c}}= \e \tilde{S} (\tau) \tilde{\boldsymbol{c}}, \label{dec2}
	\end{equation}
	where
	\begin{equation}
		\tilde{S}=T_{\tau}^{-1}(S T_{\tau} - \partial_{\tau} T_{\tau}) = \frac{\tilde{S}_0}{\tau}+\frac{\tilde{S}_{1}}{\tau-1} \label{smatr}
	\end{equation}
	with $\tilde{S}_i$ being the constant matrices. Let us write the solution of \eqref{dec2} in form
	\begin{equation}
		\tilde{\boldsymbol{c}}(\tau,\epsilon)= U_{\tau} (\tau,\underline{0}) \cdot L_{\tau}(\epsilon) \cdot \hat{\boldsymbol{c}}(\epsilon), \label{dec3}
	\end{equation}
	where
	\begin{equation}
		U_{\tau} (\tau,\underline{0}) = \lim_{\tau_0 \rightarrow 0} \biggr( \text{Pexp}\biggr[ \epsilon \int_{\tau_0}^{\tau} \bigg(\frac{\tilde{S}_0}{\tau}+\frac{\tilde{S}_{1}}{\tau-1}\bigg)d\tau \biggr] \cdot \tau_{0}^{\epsilon \tilde{S}_0} \biggr),
	\end{equation}
	$\hat{\boldsymbol{c}}(\epsilon)$ are the asymptotic coefficients of master integrals in the double asymptotics $m^2 \rightarrow 0$, then $\tau \rightarrow 0$, and $L_{\tau}(\epsilon)$ is the ``adapter'' matrix. Note that using \Libra we can obtain $U_{\tau} (\tau,\underline{0})$ in two different forms: as power series in $\e$ and as generalized series in $\tau$.
	The $\epsilon$-expansion of $U_{\tau}(\tau,\underline{0})$ is expressed in terms of the Goncharov's polylogarithms with indices $0$ and $1$ and argument $\tau\in[0,1]$. Up to weight 4, these functions can be expressed via the classical polylogarithms $\text{Li}_{k}$.

	In order to obtain the dimensional recurrence relations for $\hat{\boldsymbol{c}}$, we use the generalized expansion of $U_{\tau} (\tau,\underline{0})$ in $\tau$ and substitute  \eqref{dec3} and \eqref{trnsf2} into \eqref{drr1}. We have
	\begin{equation}
		\hat{\boldsymbol{c}}(\epsilon+1)= \widetilde{W} (\epsilon)  \hat{\boldsymbol{c}}(\epsilon), \label{dra2}
	\end{equation}
	with
	\begin{equation}
		\widetilde{W} (\epsilon)  = [T_{\tau}  U_{\tau} L_{\tau}]_{\epsilon\rightarrow\epsilon+1}^{-1} \cdot W (\tau,\epsilon) \cdot [T_{\tau} U_{\tau} L_{\tau}] . \label{dra21}
	\end{equation}
	Similar to the derivation of Eqs. \eqref{dec1}--\eqref{drr11}, we obtain $\widetilde{W} (\epsilon)$ by expanding the right-hand side of Eq. \eqref{dra21} up to sufficiently high order in $\tau$.
	Remarkably, by judicious choice of the coefficients $\hat{\boldsymbol{c}}$ we can secure the lower triangular form of the matrix $\widetilde{W}$.

	The results of the DRA method for $\hat{\boldsymbol{c}}$ are represented in terms of nested sums of the product of $\Gamma$-functions and possibly of the  $\psi^{(n)}$ function, $\psi^{(n)}(x)=\partial_x^n\ln \Gamma(x)$. It is natural to expect, and indeed is true in our case, that the results which do not contain $n$-fold sums with $n\geqslant2$ can be obtained by the direct integration of the parametric representation. However we encounter also the $2$- and $3$-fold sums, which therefore, justifies the use of the DRA method.

	In the next section we describe how we find a general solution of recurrence relations \eqref{dra2}.

	\subsection{General solution of recurrence relations}
	Thanks to the triangular form of the matrix $\widetilde{W}$, we can write the recurrence equation \eqref{dra2} for each function $\hat{c}_k(\e)$ in the form
	\begin{equation}
		\hat{c}_{k}(\epsilon + 1)=\widetilde{W}_{kk}( \epsilon ) \cdot \hat{c}_{k}( \epsilon ) + \sum_{l<k} \widetilde{W}_{kl}( \epsilon )\hat{c}_{l}. \label{dra3}
	\end{equation}
	According to Ref. \cite{lee2010space}, the solution of \eqref{dra3} can be written in the form
	\begin{equation}
		\hat{c}_{k} ( \epsilon )= \frac{\omega_{k}( \epsilon )}{\Sigma_{k}(\epsilon)} + P_k (\epsilon), \label{dra4}
	\end{equation}
	where the ``summing factor'' $\Sigma( \epsilon )$ is a particular solution of the homogeneous equation
	\begin{equation}
		\Sigma( \epsilon) = \widetilde{W}_{kk}( \epsilon )\Sigma (\epsilon +1),
	\end{equation}
	$\omega_{k}$ is an arbitrary periodic function, and $P_k$ is a particular solution of Eq. \eqref{dra3}. In applying the DRA method, we use the top-to-bottom strategy, considering the equations sequentially for increasing $k$, i.e., starting from the simplest cases.

	Let us show how to construct the general solution of Eq. \eqref{dra3} on the example of $$\hat{c}_{103}=\left[[j_{103}]_{m^0}\right]_{\tau^{-2-2\e}}$$
	which is the asymptotic coefficient in front of $m^0\tau^{-2-2\epsilon} $  of $j_{103}=\inlinegraphics[height=1cm]{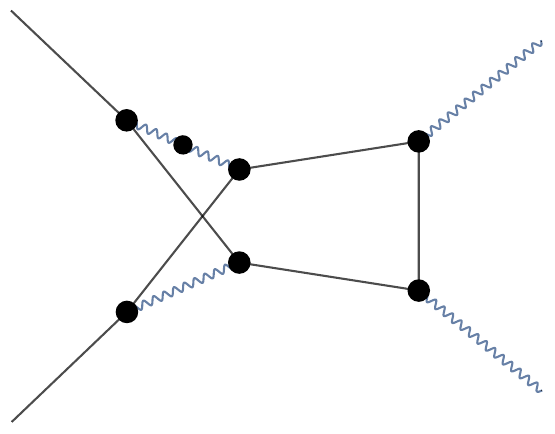}$ in the double asymptotics $m^2\to 0$,  $\tau\to 0$.
	The recurrence relation for $\hat{c}_{103}$ reads
	\begin{equation}
		\hat{c}_{103}(\epsilon + 1) =\frac{6 (2 \epsilon +3) (3 \epsilon +2) (3 \epsilon +4)}{\epsilon +2} \hat{c}_{103}(\e)
		+\sum_{l\in\{3,35,40,92\}} \widetilde{W}_{103,l}( \epsilon )\hat{c}_{l}.
		\label{ex2}
	\end{equation}
	According to the top-to-bottom strategy, we assume that at this stage we have already obtained the results for the functions  $\hat{c}_{3}$, $\hat{c}_{35}$, $\hat{c}_{40}$, $\hat{c}_{92}$ entering the right-hand side. More specifically, $\hat{c}_{3}$ is expressed via the product of $\Gamma$-functions, while  $\hat{c}_{35}$, $\hat{c}_{40}$, $\hat{c}_{92}$ are expressed via one-fold sums. The summing factor $\Sigma( \epsilon)$ satisfies
	\begin{equation}
		\Sigma( \epsilon) = \frac{6 (2 \epsilon +3) (3 \epsilon +2) (3 \epsilon +4)}{\epsilon +2}\Sigma (\epsilon +1),
	\end{equation}
	We choose the particular solution of this equation in the form  \cite{lee2010space}
	\begin{equation}
		\Sigma (\e) = -\frac{2^{-2 \epsilon -3} 3^{-3 \epsilon -\frac{1}{2}} \sin (2 \pi  \epsilon ) \Gamma \left(-\epsilon -\frac{1}{2}\right) \Gamma \left(-\epsilon -\frac{1}{3}\right)
			\Gamma \left(\frac{1}{3}-\epsilon \right)}{\pi ^{5/2} \cdot \Gamma (-\epsilon -1)}\label{Sigma}
	\end{equation}
	Using Eqs. \eqref{Sigma} and \eqref{ex2}, we obtain for the function $g(\e) = \Sigma(\e) \hat{c}_{103} (\e)$ the equation
	\begin{equation}
		g(\e +1 ) - g(\e) = r(\e) \stackrel{def}{=} \Sigma(\e + 1 )\!\!\!\!\!\! \sum_{l\in\{3,35,40,92\}} \!\!\!\!\!\!\widetilde{W}_{103,l}( \epsilon )\hat{c}_{l}. \label{ex4}
	\end{equation}
In order to write the particular solution for the inhomogeneous equation \eqref{ex4}, we decompose $r(\e)$ as
	\begin{equation}
		r(\e) = r_{+}(\e) + r_{-}( \e )+r_0(\e), \label{ex5}
	\end{equation}
	where $r_{\pm}$ decreases faster than $\frac{1}{\e}$ at $\e \rightarrow \pm\infty$ and $r_0(\e)$ does not decay sufficiently fast.
	Then the solution of \eqref{ex4} can be written in the form
	\begin{equation}
		g(\e) = \omega(\e) - \sum_{k=0} ^{\infty} r_{+}(\e  +k) + \sum_{k=0} ^{\infty} r_{-}(\e -1 -k) +g_0(\e),  \label{ex6}
	\end{equation}
	where $g_0(\e)$ is a solution of
	\begin{equation}
		g_0(\e + 1) - g_0( \e ) = r_0(\epsilon). \label{ex73}
	\end{equation}
	In principle, $r_0(\e)$ may contain (nested) sums, but we get rid of these sums by the linear change of basis of functions $\hat{c}_{k}$ as explained in Ref. \cite{lee2011application}.
	\begin{equation}
		r_{0}(\e) = R(\e)
		\psi(-\epsilon )
		\stackrel{\e\to-\infty}{=} \frac{3 \ln |\e|}{\e} + O\left( \frac{\ln^{2} |\e|}{\e^2}\right),
	\end{equation}
	where $\psi(x)=\frac{\Gamma(x)^{'}}{\Gamma(x)}$ and
	\begin{equation}
		R(\e)=\frac{27 \epsilon^3}{(\epsilon +1) (\epsilon +2) (3 \epsilon +2) (3 \epsilon +4)}.
	\end{equation}
	Unfortunately, we can not write the solution of Eq. \eqref{ex73} in a closed form. Instead, we consider an auxiliary equation
	\begin{equation}
		\tilde{g}_{0} (\e+1) - \tilde{g}_{0} (\e) = \tilde{r}_{0} (\e), \label{g0}
	\end{equation}
	and choose $\tilde{r}_0(\e)$ such that $r_{0}(\e) - \tilde{r}_{0} (\e)$ decays sufficiently fast, on one hand and Eq. \eqref{g0} has a closed-form solution, on the other hand.
	In fact, we first define the function $\tilde{g}_0$ as
	\begin{equation}
		\tilde{g}_{0} (\e) = \frac{\e}{2}R(\e) \psi(-\e)^2. \label{g01}
	\end{equation}
	Then $\tilde{r}_0(\e)=\tilde{g}_{0} (\e+1)-\tilde{g}_{0} (\e)$. Besides, we have
	\begin{equation}
		r_{0}(\e) - \tilde{r}_{0} (\e) =
		[\epsilon  R(\epsilon )-(\epsilon +1) R(\epsilon +1)]\frac{\psi(-\epsilon )^2}{2}
		+[R(\epsilon )-R(\epsilon +1)] \psi(-\epsilon)
		-\frac{R(\epsilon +1)}{2(\epsilon +1)}.
	\end{equation}
	It is easy to check that the three terms in this equation decay at $\e\to -\infty$ as $\frac{\ln^2|\e|}{\e^2}$, $\frac{\ln|\e|}{\e^2}$, and  $\frac{1}{\e^2}$, respectively.
	Then, we have
	\begin{equation}
		g_0(\e) =\tilde{g}_{0} (\e)  +\sum_{k=0} ^{\infty} [r_{0}(\e-k)-\tilde{r}_{0} (\e - k)]. \label{ex78}
	\end{equation}
	The $\epsilon$-expansion of sums in \eqref{ex78} can be calculated numerically with very high precision with the \texttt{SummerTime} package \cite{lee2016introducing}. Then the analytical result can be recognized from the numerical one using the \texttt{PSLQ} algorithm \cite{ferguson1999analysis}.
	The \texttt{SummerTime} package allows one to calculate triangular sums with a factorized summand consisting of rational or gamma functions. To reduce \eqref{ex78} to this form, we express $\psi(k-\e)$ in terms of single sums, and $[\psi(k-\e)]^2$ in terms of single and double triangular sums. For example, we use the formula
	\begin{align}
		(\psi(k-\e))^2  = 2 \sum_{l=0}^{k}  \frac{1}{l-\e-1} & \sum_{m=0}^{l} \frac{1}{m-\e-1}  - \sum_{l=0}^{k} \frac{1}{(l-\e-1)^2} 	+ \nonumber \\ &+ 2 \psi (-1-\e) \sum_{l=0}^{k} \frac{1}{l - \e -1} + (\psi(-1-\e))^2 . \label{sum1}
	\end{align}

	\subsection{Analytical properties and Mellin-Barnes representation}
	In order to construct the special solution of the dimensional recurrence relations one has to fix arbitrary periodic functions entering the general solution.	Within the DRA method, it is done using the information about analytical properties of the functions $\hat{c}_k(\e)$.
	To do this we derive Mellin-Barnes representation for those functions. Let us demonstrate our approach on the example of $\hat{c}_{103}$.

	First, we write the parametric representation of the master integral $j_{103}$:
	\begin{equation}
		j_{103}=\texttt{gg}_2(1,1,1,1,1,1,2,0,0|m^2,\tau,d)=\frac{e^{2\epsilon \gamma}\Gamma(2-\e)}{\Gamma(-2-3\e)}\int_{\mathbb{R}_+^7}dx_1\ldots dx_7 \,\frac{x_7}{G^{2-\e}}, \label{repr1}
	\end{equation}
	where
	\begin{multline}
		G= \left(1-\tau -m^2\right)x_1 x_3 x_5
		+ \left(m^2+\tau \right)x_3 x_6 x_7
		-x_2 x_5 x_6-x_1 x_4 x_7-x_2 x_4 x_{1567}\\
		-m^2 \left(x_6 x_7 x_{15}+x_1 x_5 x_{67}+x_{17} x_{56} x_{234}\right)
		+\left(x_{16} x_{57}+x_{234} x_{1567}\right) \left(m^2 x_{12345}+1\right),
	\end{multline}
	$x_{ij\ldots k}=x_i+x_j+\ldots x_k$. In order to get a representation for the asymptotic coefficient $\hat{c}_{103} = \left[[j_{103}]_{m^0}\right]_{\tau^{-2-2\e}}$, we use the expansion by regions method. Using the \texttt{asy} \cite{pak2011geometric} package, we find the integration regions that contribute to this asymptotic coefficient. First, we consider $m^2$ as small parameter and find one region corresponding to the naive limit $m\to 0$. Thus, $[j_{103}]_{m^0}$ is defined by the same Eq. \eqref{repr1} with $G$ replaced by $\widetilde{G} = \left.G\right|_{m=0}$. Then we consider the small-$\tau$ asymptotics of $[j_{103}]_{m^0}$ and find two regions
	\begin{align}
		\mathrm{I.}&\ x_1\sim x_2\sim x_3\sim x_6\sim x_7\sim \tau^{-1},\ x_4\sim x_5 \sim 1,
		\\
		\mathrm{II.}&\ x_3\sim x_4\sim x_5 \sim x_6\sim x_7\sim \tau^{-1},\  x_1\sim x_2\sim 1.
	\end{align}
	It appears that, in order to separate the contribution of these regions, we need to use the analytical regularization. We do this by introducing the factor $x_1^{a}$ in  the integrand. Then, using the Mellin-Barnes parametrization, we obtain the contributions of the two regions in the form
	\begin{align}
		\hat{c}_{103,\mathrm{I}} & =- \frac{e^{2\epsilon \gamma}\Gamma (-a-\epsilon,2+a+2\epsilon)}{\Gamma (-2 - a -3\epsilon,-1-a-2\epsilon)}  \int \frac{dz_{1} dz_{2}}{(2 \pi i)^2} \,\,
		e^{i \pi  z_1}
		\frac{\Gamma (z_1,z_2,a+z_1+z_2+\epsilon )}{\Gamma (-z_1-z_2-\epsilon +1)}
		\nonumber \\
		&\times
		\Gamma^{2} \left(1-z_1\right) \Gamma (z_1-a-2 \epsilon -2,-z_2-\epsilon,-a-z_1-z_2-2 \epsilon -1,-z_1-z_2-\epsilon ) ,
		\nonumber\\
		\hat{c}_{103,\mathrm{II}} & =-\frac{e^{2\epsilon \gamma}\Gamma (-1-\epsilon,2+2\epsilon)}{\Gamma (-2-a-3\epsilon,-1-a -2\epsilon)}  \int \frac{dz_{1} dz_{2}}{(2 \pi i)^2}  \,\,  e^{i \pi z_1}
		\frac{\Gamma (z_1,z_2,z_1 + z_2 +\epsilon-a,1-z_1) }{\Gamma(1 - z_1 - z_2 -\epsilon)}
		\nonumber \\
		& \times   \Gamma (- z_1 - z_2 -\epsilon,1+a-z_1,-2-a+z_1 -2\epsilon,-z_1 - z_2 -2\epsilon,- z_2 -\epsilon), \label{mb3}
	\end{align}
	where $\Gamma(a_1,a_2,\ldots,a_n)=\Gamma(a_1)\Gamma(a_2)\ldots\Gamma(a_n)$. These Mellin-Barnes integrals are well-defined when the integration contours over $z_1$ and $z_2$ and the parameters $\epsilon$ and $a$ are chosen in such a way that all arguments of $\Gamma$-functions in the numerator of the integrand have positive real parts. Using the \texttt{MB} package\footnote{We thank A. Pikelner for providing a patch for this package able to use the \texttt{C} integration libraries.} \cite{czakon2006automatized}, we can analytically continue Eqs. \eqref{mb3} to any region of the parameters $a$ and $\e$. In particular, it is easy to establish that the expansion of each of $\hat{c}_{103,\mathrm{I}}$ and $\hat{c}_{103,\mathrm{II}}$ at $a=0$ starts from $a^{-1}$, while their sum has a finite limit $a\to 0$.

	Using these representations we find out that the asymptotic coefficient has poles of the second and third order at the points $\e=-1/2$ and $\e=-1$, respectively, while being analytical everywhere else on the stripe $S=\left\lbrace \e\in \mathbb{C} \; | \; -3/2 < \Re \e < -1/2 \right\rbrace $. Consequently, using the explicit form \eqref{Sigma}, we establish that  $\Sigma (\e) \cdot \hat{c}_{103}(\e)$ has poles of the second and first order at $\e=-1/2$ and $\e=-1$, respectively, and analytical everywhere  else on the stripe $S$. Performing numerical integration with the \texttt{MB} package we obtain
	\begin{align}
		e^{-2\epsilon \gamma}\Sigma \cdot \hat{c}_{103} & = -\frac{1.50000\ldots}{ (\e+1/2)^2} - \frac{5.07944\ldots + i\cdot 9.42478\ldots}{\e+1/2} + O\left((\e+1/2)^0\right) , \nonumber \\
		e^{-2\epsilon \gamma}\Sigma \cdot \hat{c}_{103} & = - \frac{i\cdot 4.71239\ldots}{\epsilon+1} + 22.2066\ldots + O(\epsilon+1) , \label{mb6}
	\end{align}
	where the numerical constants were calculated with the precision of $50$ digits. Then, using the \texttt{PSLQ} and educated guess about the constants, we obtain
	\begin{align}
		e^{-2\epsilon \gamma}\Sigma \cdot \hat{c}_{103} & =  -\frac{3}{2 (\epsilon+1/2)^2} - \frac{3 + 3 \ln2 + 3 i\pi}{\epsilon+1/2}  + O\left((\e+1/2)^0\right), \nonumber \\
		e^{-2\epsilon \gamma}\Sigma \cdot \hat{c}_{103} & =  - \frac{ 3  i\pi}{2(\epsilon+1)} + \frac{9 \pi^2}{4} + O(\epsilon+1), \label{mb61}
	\end{align}
	These expansions, together with the analyticity in all other points of $S$, allow us to fix the special solution of dimensional recurrence relation \eqref{ex2}. After this we can calculate the expansion of $\hat{c}_{103}$ in any point up to any order with arbitrary precision. In particular, we obtain
	\begin{align}
		\hat{c}_{103}(\e) \overset{1000}{=}  &-\frac{2}{\epsilon^4} + \frac{-5 + 2 i\pi}{\epsilon^3}+ \frac{5-\frac{9 \zeta_{2}}{2} + 15 i\pi}{\epsilon^2} +  \frac{1-39 \zeta_{2} + \frac{65 \zeta_{3}}{6} + i\pi (\frac{29}{3}-29 \zeta_{2})}{\epsilon} - \nonumber \\ & - 10 -60\zeta_{2} -\frac{2\zeta_{3}}{3} + \frac{147\zeta_{2} ^{2}}{5}  + i\pi \left(\frac{31}{9} - 123\zeta_{2} - \frac{271\zeta_{3}}{3}\right) + \dots + O(\epsilon^9). \label{mb7}
	\end{align}

	In a similar manner, we obtain the expressions for all asymptotic coefficients $\hat{\boldsymbol{c}} (\epsilon)$ in terms of triangular sums exact in $\e$. Using these expressions we obtain their high-precision expansions at $\e=0$ and recognize the coefficients in terms (alternating) multiple zeta values.

	\subsection{Uniform transcendental basis}
	We calculate the asymptotic coefficients $\hat{\boldsymbol{c}} (\epsilon)$ \eqref{dec3} up to the constants of transcendental weight 8.
	This depth of $\e$-expansion appears to be sufficient to find the uniform transcendental (UT) basis for the boundary constants
	\begin{equation}
		\hat{\boldsymbol{C}}(\e)=L_{\tau}(\e) \hat{\boldsymbol{c}} (\epsilon)\,.
	\end{equation}
	Note that the canonical basis $\tilde{\boldsymbol{c}}$ satisfying Eq. \eqref{dec2} is defined up to an arbitrary constant transformation $T_{ut} (\e)$ that commutes with the matrix $\tilde{S}(\tau)$,
	\begin{equation}
		\tilde{S} (\tau) T_{ut} (\e) = T_{ut} (\e) \tilde{S} (\tau) . \label{ut0}
	\end{equation}
	We want to construct $T_{ut} (\e)$ such that
	\begin{equation}
		\widetilde{\boldsymbol{C}}(\tau,\e)\stackrel{def}{=}  T_{ut}^{-1}(\e)\tilde{\boldsymbol{c}}(\tau,\e)
		=U_{\tau} (\tau,\underline{0})  T_{ut}^{-1}(\e) L_{\tau}(\epsilon)  \hat{\boldsymbol{c}}(\epsilon).
	\end{equation}
	has a UT expansion. Here we used that $T_{ut}^{-1}U_{\tau} (\tau,\underline{0})  =U_{\tau} (\tau,\underline{0})  T_{ut}^{-1}$ thanks to Eq. \eqref{ut0}.

	The evolution operator $U_{\tau} (\tau,\underline{0})$ has a UT expansion by construction. Therefore, in order to find the transcendental basis, we have to choose $T_{ut}$ such that
	\begin{equation}
		\hat{\boldsymbol{C}}_{ut}(\e) = T_{ut}^{-1}\hat{\boldsymbol{C}}(\e) = T_{ut}^{-1}L_\tau\hat{\boldsymbol{c}}
	\end{equation}
	is uniform transcendental.

	Note that there is always at least a one-parametric family of matrices satisfying Eq. \eqref{ut0}, namely $T_{ut}=f(\e) I$, where $f(\e)$ is a rational function and $I$ is an identity matrix. If we are lucky enough, we can obtain the UT basis by finding the appropriate rational function $f(\e)$. Along these lines we search for $\hat{C}_{ut,103}(\e)$ in the form
	\begin{equation}
		\hat{C}_{ut,103}(\e)= f(\e)\hat{C}_{103}(\e)
	\end{equation}
	and find that the choice $f(\e)=\e(1-2\e)(1-3\e)(2-3\e)$ entails the UT expansion of $\hat{C}_{ut,103}(\e)$.
	However, in many other cases, such an ansatz fails. This is connected with the fact that the family of matrices satisfying Eq. \eqref{ut0} has a lot of free parameters. It can be easily established by using the standard \Libra procedure \texttt{FactorOut}. It appears that the matrix $T_{ut}$ contains as many as $220$ free parameters.
	Therefore, we follow the top-to-bottom strategy to fix these parameters. For a specific constant $\hat{C}_{ut,k}(\e)$, we try to nullify as many of yet unfixed parameters entering this constant as possible so that the remaining parameters were still sufficient to ensure the UT expansion.

	Of course, the described approach is intrinsically heuristic and requires a cross check. To this purpose we calculate the candidate UT basis up to $\e^{10}$ and observe the UT expansion.

	\subsection{Comparison with known results and numerical check}

	We perform several cross checks of our results. First, some of the asymptotic coefficients $c_i$ correspond to the `naive' limit $m^2\to 0$, i.e. the values of the corresponding massless integrals. For those coefficients we find a perfect agreement of our results with analytical results of Refs. \cite{smirnov2000analytical,tausk1999non} as well as with the numerical results obtained using \texttt{FIESTA5} \cite{smirnov2022fiesta5}.

	We have also compared the Frobenius expansions of the massive integrals obtained in this work with the results for those massive integrals obtained with \texttt{FIESTA5}.

	Besides, we obtain the numerical results for the massive master integrals at $d=6-2\e$ using \texttt{FIESTA5}\footnote{Note that \texttt{FIESTA5} is much more efficient in calculating the most complicated master integrals at $d=6-2\e$ than at $d=4-2\e$.} for sufficiently small $m^2$ (see below).

	Our results for $d=6-2\e$ master integrals
	can be obtained by combining \eqref{transf1} and \eqref{drr0},
	\begin{eqnarray}
		\left.\boldsymbol{j}\right|_{d=6-2\e} = \boldsymbol{j}(\e-1) = M_{\e}^{-1}T_{m} \boldsymbol{J}.
	\end{eqnarray}

	It should be noted that expansion of $M_{\e}^{-1}T_{m}$ starts from $m^0$, therefore when multiplying the Frobenius expansion of $\boldsymbol{J}$ by this matrix, we do not lose the higher orders in $m^2$ and obtain the expansion for $\left.\boldsymbol{j}\right|_{d=6-2\e} $ up to $m^{12}$. This depth of expansion was sufficient to compare with the \texttt{FIESTA5} results. At $d=6-2\epsilon$, $m^2=\frac{1}{20}$ and $\tau = -t = \frac{6}{10}$, the numerical and analytical results differ by no more than $10\%$. The $10\%$ deviation for some integrals in the upper sectors is due to the slow convergence of the series in $m^2$. By choosing a somewhat smaller value, $m^2=\frac{1}{50}$, we find the agreement within $0.5 \%$.

	\section{Results and ancillary files}\label{sec:files}
	Since the obtained expressions are too cumbersome to be presented in the text, we attach all the results in a computer-readable form.

	Let us represent the master integrals in $d=6-2\e$ in a form:
	\begin{equation}
		\left.\boldsymbol{j}\right|_{d=6-2\e}
		= (m^2)^D \cdot \biggr[ \sum_{i=1}^{7}(m^2)^{\lambda_{i}}\sum_{n=0}^{6}\sum_{k=0}^{K_{i}}  B_{i,n,k} \cdot m^{2n}\cdot \frac{ \ln^{k} m^2}{k!} \biggr] \cdot \widetilde{\boldsymbol{C}}(\tau,\e), \label{file1}
	\end{equation}
	where $B_{i,n,k}$ are the matrix coefficients exact in $\e$ and $\tau=-t$ (recall that we set $s=1$), $\tilde{\boldsymbol{C}}$ is a vector of functions in the form of a UT $\e$-series with coefficients expressed in terms of alternating multiple zeta values and Goncharov's polylogarithms of argument $\tau=-t$ with indexes $0$ and $1$. The matrix $D$ has the form
		$D=\mathrm{diag}(o_1,\ldots,o_{165})$,
	where $m^{2o_l}$ is the leading small-mass expansion term of $\left.j_l\right|_{d=6-2\e}$.

	We attach the following files:
	\begin{itemize}
		\item \texttt{ggs.m} contains the definitions of \LiteRed bases.
		\item \texttt{MIs.m} contains the definitions of master integrals in terms of integrand in the momentum space. Note that there are 4 master integrals with irreducible numerators.
		\item  \texttt{BtildeC.m} contains a depth-3 list of $B_{i,n,k}\widetilde{\boldsymbol{C}}$. This file is for convenience only. It is used in the example notebook \texttt{EvaluateMasters.nb} (see below).
		\item \texttt{Ddiag.m} contains the diagonal $(o_1,\ldots,o_{165})$ of the matrix $D$.
		\item \texttt{EvaluateMasters.nb} is the example \texttt{Mathematica} file with numerical calculation.
	\end{itemize}
These files together with extended data listed below are available on the github repository \url{https://github.com/VStotsky/ee2gamma_nnlo/}.

	The repository also contains
	\begin{itemize}
	\item \texttt{B.m} --- a depth-3 list of matrix coefficients $B_{i,n,k}$.
	\item \texttt{tildeC.m} --- a list of  $\widetilde{\boldsymbol{C}}(\tau,\e)$ in the form of UT $\epsilon$-series up to $\e^8$.
	\item \texttt{ShiftDimension.m} --- the rules for expressing master integrals in $d=4-2\e$ through master integrals in $d=6-2\e$.
\end{itemize}
	\section{Conclusion}
	In the present paper, we have calculated the two-loop massive master integrals for $e^{+}e^{-}\rightarrow 2\gamma$ in the high energy approximation. The results were obtained in terms of generalized series up to $\left({m}/{\sqrt{s}}\right)^{12}$ with coefficients expressed via  Goncharov's polylogarythms up to weight $8$. A uniformly transcendental basis of functions defining the coefficients of these series was also obtained.	The calculation was carried out using modern  multi-loop calculation methods.

	In the future, the results of this work will be used to calculate the differential cross section  of $e^{+}e^{-}\rightarrow 2\gamma$ with accuracy $O(\alpha^4)$, as well as to test the massification formulae.
	\acknowledgments
	We thank A. Pikelner for providing a patch for \texttt{MB} package able to use the \texttt{C} integration libraries. We are grateful to V.S. Fadin for the interest to this work and helpful discussions. The work is supported by the Russian Science Foundation, grant number 20-12-00205.

	\bibliographystyle{JHEP}
	\bibliography{2lmassive}
\end{document}